\documentclass{article}

\usepackage{arxiv}

\usepackage[utf8]{inputenc} 
\usepackage[T1]{fontenc}    
\usepackage{hyperref}       
\usepackage{url}            
\usepackage{booktabs}       
\usepackage{amsfonts}       
\usepackage{nicefrac}       
\usepackage{microtype}      
\usepackage{lipsum}		
\usepackage{graphicx}
\usepackage{natbib}
\usepackage{doi}

\usepackage[utf8]{inputenc} 
\usepackage{textcomp}       
\usepackage{algorithm}
\usepackage{algpseudocode}
\usepackage{tabularx}
\usepackage{amsmath}

\title{Data assimilation approach for addressing imperfections in people flow measurement techniques using particle filter}

\author{
  Ryo Murata \\
  Graduate School of Engineering, \\
  Department of Technology Management for Innovation \\
  The University of Tokyo \\
   \And
  Kenji Tanaka \\
  Graduate School of Engineering, \\
  Department of Technology Management for Innovation \\
  The University of Tokyo \\
}

\renewcommand{\shorttitle}{}

\hypersetup{
pdftitle={A template for the arxiv style},
pdfsubject={q-bio.NC, q-bio.QM},
pdfauthor={David S.~Hippocampus, Elias D.~Striatum},
pdfkeywords={First keyword, Second keyword, More},
}

\begin{document}
\maketitle

\begin{abstract}
Understanding and predicting people flow in urban areas is useful for decision-making in urban planning and marketing strategies. Traditional methods for understanding people flow can be divided into measurement-based approaches and simulation-based approaches. Measurement-based approaches have the advantage of directly capturing actual people flow, but they face the challenge of data imperfection. On the other hand, simulations can obtain complete data on a computer, but they only consider some of the factors determining human behavior, leading to a divergence from actual people flow. Both measurement and simulation methods have unresolved issues, and combining the two can complementarily overcome them. This paper proposes a method that applies data assimilation, a fusion technique of measurement and simulation, to agent-based simulation. Data assimilation combines the advantages of both measurement and simulation, contributing to the creation of an environment that can reflect real people flow while acquiring richer data. The paper verifies the effectiveness of the proposed method in a virtual environment and demonstrates the potential of data assimilation to compensate for the three types of imperfection in people flow measurement techniques. These findings can serve as guidelines for supplementing sparse measurement data in physical environments.
\end{abstract}


\section{Introduction}
Recent advancements in mobile devices have made it possible to observe people's behaviors in detail. For example, GPS in smartphones can track the two-dimensional movements of people across wide areas, like entire cities. Additionally, by using Wi-Fi or Bluetooth signals, it's possible to track three-dimensional movements in areas where GPS has limited reach, such as inside buildings or underground. The observation of human behavior in urban areas is becoming increasingly capable of covering our range of activities. This wealth of information obtained from numerous sensors can be useful in decision-making for urban planning(\cite{RefWorks:RefID:36-li2019experimental}) and marketing  strategies(\cite{RefWorks:RefID:30-erevelles2015big}).

However, current measurement techniques have limitations. For example, traditional camera-based measurement can only track movements within the camera's field of view, making wide-area measurement difficult. Bluetooth/Wi-Fi-based measurement are limited to people who have installed specific apps, hindering comprehensive measurement. These imperfections in people flow measurement methods limit the acquisition of information for understanding human behavior or lead to incorrect decisions due to biased sampling(\cite{RefWorks:RefID:31-malinovskiy2012analysis}).

Alongside measurements, people flow simulations have been used to understand human behavior. In simulations, models are created based on variables influencing human behavior, allowing for computer-based pseudo-verification for understanding phenomena and prediction. Unlike measurements, simulations can acquire complete data as people's movements are recorded on the computer. Especially, advancements in computing power have enabled the execution of methods like agent-based simulations, which account for the diverse behavioral characteristics of each individual. Such models can describe not just crowd behavior during events or evacuations but also everyday behaviors like city roaming(\cite{RefWorks:RefID:32-malleson2022agent-based}), and purchasing activities(\cite{RefWorks:RefID:60-ghazavi2016formulation}). This expands the scope of simulations to include everyday scenarios, not just specific situations where crowds form.

However, simulations alone can diverge from actual people flow, making it challenging to extract practical insight(\cite{RefWorks:RefID:50-wijermans2016landscape}). This is because human flow models extract only a portion of the factors that influence people's behaviors, while disregarding other factors. Especially in scenarios with diversity in individual destinations, such as roaming behavior, the difference between simulation and actual people flows can become significant.

To address the imperfections of people flow measurement and the divergence from real environments in simulation, the technique of data assimilation, rooted in earth sciences, can be employed. Data assimilation is a hybrid method of measurement and simulation, incorporating observational data into simulation to make accurate state estimations. In earth sciences, this method has been used to reasonably estimate hard-to-observe physical quantities from other observational data. Applying data assimilation to the simulation of people flow can complement the limitations of measurement techniques. This approach allows for the acquisition of comprehensive data that accurately reflects actual people flow.

This paper proposes a data assimilation method that combines agent-based simulation and particle filtering to supplement the imperfection in people flow measurement data. First, we organize the current major measurement techniques and describes the imperfections they are facing. Then, we explain how the proposed method can complement these imperfections. Finally, to validate our proposed method, we conduct an experiment in a virtual commercial facility, estimating the overall people flow from the individual store's people inflow count data and comparing these estimates with pseudo-observational data.

The contributions of this paper are as follows:
\begin{itemize}
  \item Organizing the main imperfections in current people flow measurement techniques and classifying them into three categories.
  \item Proposing a fusion method of measurement and simulation to supplement the imperfections in people flow measurement techniques.
  \item Conducting experiments in a virtual environment and demonstrating the effectiveness of the proposed method.
\end{itemize}

\section{Related Work}
\subsection{Measurement techniques for  people flow tracking}
Techniques for measuring people flow are divided into vision-based and non-vision-based methods(\cite{RefWorks:RefID:33-singh2020crowd}). Vision-based methods mainly use cameras and typically involve recognizing people through supervised learning. Cameras can estimate attributes like age and gender from image data and associate them with the trajectories. However, cameras cannot track people flow outside their field of view, making them unsuitable for wide-area measurement such as Origin-Destination (OD) data. Additionally, camera-based measurement often involves personal information, necessitating cautious management to ensure privacy protection.

Non-vision-based techniques, which have emerged as alternatives to vision-based methods, enable wide-area measurement and are more privacy-conscious. Typical examples include GPS, Bluetooth/Wi-Fi, LiDAR, and RFID(\cite{RefWorks:RefID:33-singh2020crowd}). 

GPS tracks continuous trajectories based on satellite-device communication and can track wide-area people movements. GPS data is frequently linked with attribute information, enabling the measurement of people flow categorized by attributes. However, GPS cannot measure indoors or underground where satellite signals are unavailable(\cite{RefWorks:RefID:36-li2019experimental}).

Bluetooth/Wi-Fi primarily detects signals emitted from smartphones using beacons. By installing beacons at multiple locations, it's possible to identify points visited by targets over a wide area, making it suitable for understanding OD information. However, this method is limited to measuring only devices that have Wi-Fi/Bluetooth turned on and specific applications installed, which restricts the scope of measurement targets.

LiDAR detects the passage of people by acquiring the three-dimensional shape of an object through laser illumination. LiDAR is privacy-conscious measurement techniques because it detects people by capturing point cloud data that does not identify individuals. However, it is not suitable for tracking wide-area movements such as  OD information.

RFID detects people movements by reading RFID tags. By installing readers at multiple locations, it can measure wide-area movements of people. However, RFID measurement can only track those equipped with RFID tags, which restricts the scope of measurement targets.

The characteristics of these representative sensors are summarized in Table\ref{sensors}. 

Each sensor exhibits particular advantages and limitations, resulting in an unavoidable imperfection of measurements in capturing comprehensive information.
\begin{table}
\centering
\caption{The list of measurable and non-measurable quantities by typical people flow measurement techniques.}
\label{sensors}
\begin{tabularx}{0.9\textwidth}{l|XXXXX} 
               & Count & OD   & Attribute & Indoor & Coverage \\ \hline
GNSS           & No   & Yes    & Yes       & No     & No       \\
Camera         & Yes   & No     & Yes       & Yes    & Yes      \\
Bluetooth/Wifi & Yes    & Yes    & No        & Yes    & No       \\
LiDAR          & Yes   & No     & No        & Yes    & Yes      \\
RFID           & Yes    & Yes    & No        & Yes    & No      
\end{tabularx}
\end{table}


\subsection{Agent-based model for people flow simulation}
Agent-based simulation is a bottom-up approach for revealing the mechanisms of macro phenomena like people flow by defining the behavior rules and interactions of agents. Even with simple behavioral rules for each agent, the accumulation of probabilistic behavior choices and interactions allows for simulating complex systems.

People flow models are broadly divided into macro models, which view people flow as a fluid to describe overall behavior, and micro models, which describe the behavior of individual people(\cite{RefWorks:RefID:48-martinez-gil2017modeling}). The agent-based model is classified as a micro model, capable of describing people flow considering differences in individual behavioral characteristics.

Agent-based models often describe people flow in specific scenarios such as evacuation(\cite{RefWorks:RefID:34-shi2009agent-based}) and infection spread(\cite{RefWorks:RefID:69-lorig2021agent-based}). In such cases, models like the Social Force Model(\cite{RefWorks:RefID:64-helbing1995social}), which focus on behaviors that maintain a certain distance while moving towards a destination, are commonly used. While these models are effective in scenarios involving high-density crowds and limited destinations, they are not as well-suited for situations with a variety of destinations and paths, such as roaming behaviors.

Examples of roaming behavior models include models describing purchasing behavior(\cite{RefWorks:RefID:60-ghazavi2016formulation}, \cite{RefWorks:RefID:61-dijkstra2014modeling}) and everyday movements in urban areas(\cite{RefWorks:RefID:49-crols2019quantifying}). This paper proposes a data assimilation method applicable to roaming behavior models with significant differences in individual behaviors. Since roaming behavior is diverse, it is challenging to accurately describe its dynamics solely with models. Therefore, by combining the data assimilation method with agent-based simulation, we aim to improve the accuracy of people flow estimation. 

\subsection{Agent-based simulations with data assimilation }
Applying data assimilation to agent-based simulation enables high-accuracy simulation of real-world people flow. The Kalman Filter, a standard method in data assimilation, is applicable to linear, Gaussian state transition models(\cite{RefWorks:RefID:67-welch1995introduction}). However, for rule-based models like agent-based models that determine individual state transitions, an analytical state description is challenging, so the Kalman Filter itself is not directly usable. Extensions of the Kalman Filter include the ensemble Kalman filter(\cite{RefWorks:RefID:44-togashi2018ensemble}) and the unscented Kalman filter(\cite{RefWorks:RefID:63-clay2020real-time}), applicable to nonlinear models. These methods can perform data assimilation for nonlinear state transition models with relatively low computational cost, but they require Gaussian assumptions in state transition models, limiting their application.

The particle filter is applicable to nonlinear and non-Gaussian state transition models. Previous studies that have used particle filters in agent-based simulations include those that improve route selection parameters in evacuation behavior(\cite{RefWorks:RefID:45-makinoshima2022crowd}) or those integrating data from multiple sensors at live events(\cite{RefWorks:RefID:66-2019there?}). These are examples of describing the flow of people when crowds move in a unified manner.

For people flows with significant differences in individual behaviors, such as roaming, examples include simulations in a train station(\cite{RefWorks:RefID:43-ternes2021data}, \cite{RefWorks:RefID:41-malleson2019simulating}) or a single floor of a building(\cite{RefWorks:RefID:42-wang2015data}). These focus on real-time operation, targeting parts of facilities or very few individuals. While the method proposed in this study also has potential for real-time estimation applications, we use particle filters to complement the imperfections of people flow measurement techniques. Therefore, the study proposes a data assimilation method for scenarios where a large number of people roam in extensive spaces like entire facilities.

\section{Data Assimilation Framework}
\subsection{Particle filter}
The particle filter is a data assimilation method that approximates the system model using Monte Carlo sampling(\cite{RefWorks:RefID:62-djuric2003particle}). It constructs a distribution based on a group of samples extracted from models, making it applicable to non-linear and non-Gaussian state-space models. This flexibility allows it to be applied to complex phenomena resulting from interactions between agents, as seen in agent-based simulations.

The primary goal of a particle filter is to estimate the current state of the system based on accumulated observational data up to the present. Let's consider the situation where the state at time $t-1$ is estimated based on observational data up to time $t-1$. The state at time $t-1$ can be represented as follows:

\begin{figure}[b]
\centering
\includegraphics[width=0.9\textwidth]{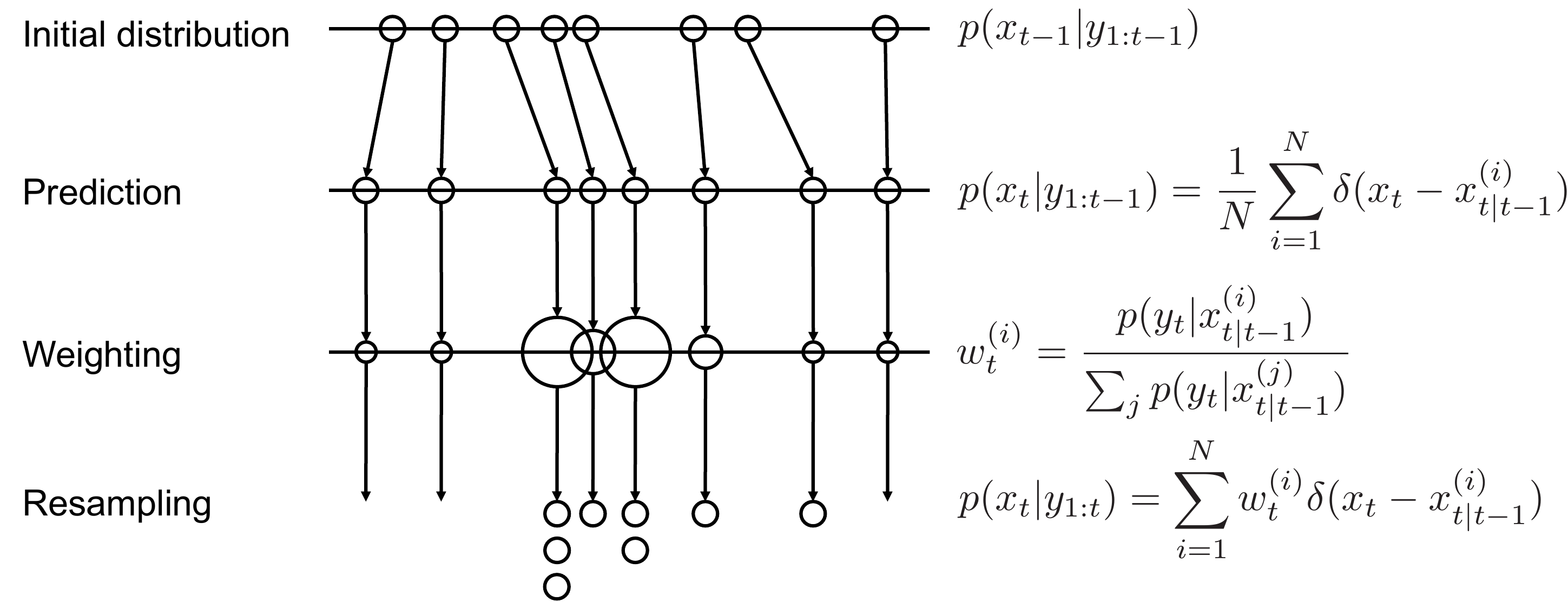}
\caption{The basic process of particle filtering. Initially, a large number of particles are generated for each entity, with the model projecting the state for the next step. The weight of each particle is determined by the likelihood calculated from observational data. A resampling process based on this likelihood is then conducted to establish the distribution of the predicted state for the next step.}
\label{particle}
\end{figure}

\begin{eqnarray}
x_{t-1} \sim p(x_{t-1} | y_{1:t-1})
\end{eqnarray}

where $x_{t-1}$ represents the state at time $t-1$ and $y_{1:t-1}$ represents the measurement from time $0$ to $t$.  
Next, the state at time $t$ is estimated based on the observations at time $t-1$ as follows:

\begin{eqnarray}
p(x_t | y_{1:t-1}) = \frac{1}{N} \sum_{i=1}^N \delta(x_t - x_{t|t-1}^{(i)})
\end{eqnarray}

where $\delta$ represents the Dirac delta function, and $N$ is the total number of particles. $x_{t|t-1}^{(i)}$ is the state of the i-th particle estimated by the transition model. This expression implies that the state at time $t$ is approximated by an ensemble of $N$ particles based on the observational data up to time $t-1$.

Then, the state estimation at time $t$ is performed based on the observations obtained at time $t$. This can be represented as follows:
\begin{eqnarray}
p(x_t | y_{1:t}) = \sum_{i=1}^N w_t^{(i)} \delta(x_t - x_{t|t-1}^{(i)})
\end{eqnarray}

where $w_t^{(i)}$ represents the weight of particle $i$. During the resampling process, particles are generated in accordance with these weights. The weight $w_t^{(i)}$ is calculated as follows:

\begin{eqnarray}
w_t^{(i)} = \frac{p(y_t | x_{t|t-1}^{(i)})}{\sum_j p(y_t | x_{t|t-1}^{(j)})}
\end{eqnarray}

where $p(y_t | x_{t|t-1}^{(i)})$ represents the probability distribution of the observational value given the state, which is known as the likelihood. Likelihood indicates how well each particle explains the observational data. By resampling particles based on likelihood, particles closer to the observational data are preferentially selected.

Figure\ref{particle} shows a  diagram of a particle filter process. By state prediction through the model and resampling weighted by likelihood, the simulation can estimate the system state accurately.

\subsection{Integrating agent-based simulation with particle filters}
When incorporating a particle filter into agent-based simulation, the key considerations are 1) defining the state variables and 2) defining the observation variables and likelihood. The state variables should be chosen to characterize roaming behavior. Observation variables should be defined as those that provide valuable information for estimating actual state. Likelihood should be defined in a way that reduces the uncertainty of the simulation based on the observation variables. By considering these aspects, we can incorporate real-world roaming tendencies into agent-based simulation via particle filtering.

Firstly, the state variable is defined as each agent's location. This is because comprehensive measurement of agent's location is difficult, and the information is important for characterizing the roaming behavior. To estimate the next step locations of agents, particles are generated for each agent based on models. These particles stochastically determine the destination, influenced by other agents. 


Observation variables are chosen as the inflow count data at each point. This is because by sequentially measuring the inflow data at each location, it is possible to calculate the likelihood based on the real-world visitation tendencies. In other words, by using inflow count data to determine which stores are more or less likely to be visited, it is possible to align the roaming tendencies of the agents more closely with those in the real world. This allows for the transition of agents based on transition models and some measured data to reflect real people flow tendencies.

Based on these methods, the simulation proceeds as follows:
\begin{enumerate}
    \item Select an agent and perform state transitions for the number of particles.
    \item Calculate the likelihood of each particle.
    \item Normalize the likelihoods to convert them into a probability distribution.
    \item Resample particles based on the probability distribution.
    \item Select one particle from the resampled ones as the agent's location for the next step.
\end{enumerate}

Performing these five steps for all agents bridges the gap between the simulation and real-world people flow.

\begin{figure}[b]
\centering
\includegraphics[width=0.8\textwidth]{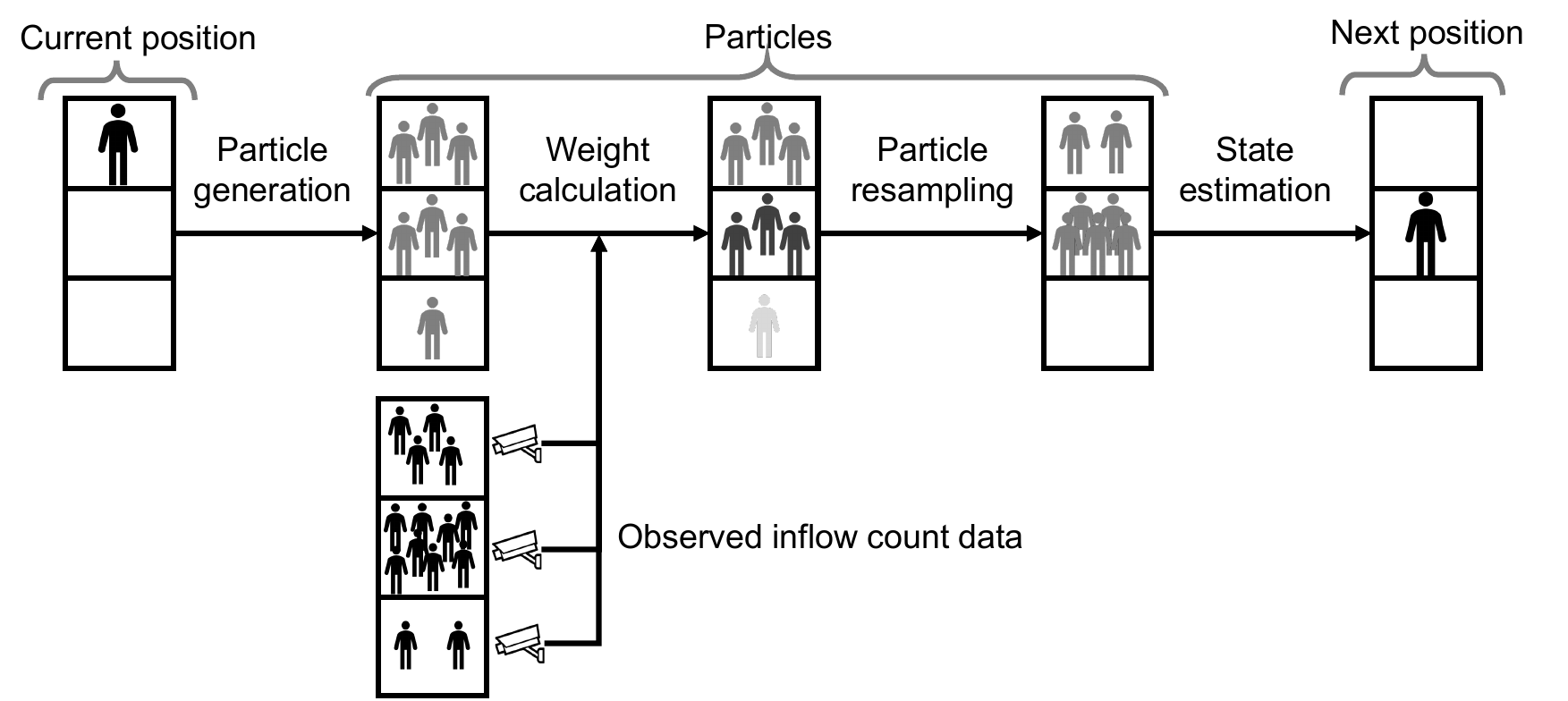}
\caption{The application of particle filtering in agent-based simulation. Particles are generated for each agent, and the model predicts their next location. Each particle's likelihood is determined based on anonymous inflow count data, and the next transition location is determined through resampling based on this likelihood.}
\label{particle_agent}
\end{figure}

\subsection{Addressing the imperfection of human flow data}
As mentioned in related works, sensor imperfection can be broadly classified into three categories. Below, we discuss guidelines for data assimilation to address these three types of imperfection.

\subsubsection{Missing travel data}
When the measurement range of sensors is limited, it can lead to the absence of comprehensive travel data, including detailed point-to-point transition information. This issue is particularly prevalent with devices like cameras or LiDAR, which hinders the collection of complete OD data across entire areas.

To address the travel data missing within agent-based simulations, we propose a method that leverages local measurement data. The approach involves determining the likelihood of an agent's next destination based on visitation tendencies. As outlined in the preceding section, this method utilizes inflow count data from each location to gauge the visitation tendencies within the space, and then incorporates these tendencies into the simulation through a particle filter. 


\subsubsection{Lack of attribute data}
Lack of attribute data refers to the inability to link attribute information with travel data. This limitation is common in many sensors, with the exception of GPS devices. In roaming behavior, attribute information, which clarifies the behavioral characteristics for each attribute, is essential for conducting a detailed analysis.

In agent-based simulations, linking attribute information with transition data can be achieved by estimating the transition of agents based on observational data that includes the attribute information. Currently, cameras serve as the primary means of capturing attribute information in both indoor and outdoor settings. However, they fall short in tracking comprehensive movements, like transitions between points. Given the limitations of current measurement techniques, it is necessary to estimate transition data from local people flow data with attribute information gathered through camera observations.

For estimating transition data with attributes using cameras, it is effective to gather inflow count data for each attribute and estimate the transition tendencies. This approach applies the method discussed in the previous subsection to each attribute identified by cameras. Its main benefit lies in reducing prediction uncertainty. In the previously mentioned method, inflow count data does not specify individual agents, so predictions are based solely on the general movement tendencies of all agents. However, this could lead to significant errors in predictions if there are variations in behavioral characteristics among different attribute groups. By using cameras to identify attributes and estimate transitions for each one, it's possible to account for these differences in behavior. This approach leads to improved accuracy in predicting overall transition tendencies.

\subsubsection{Limited measurement scope}
A limited measurement scope restricts the ability to observe everyone in a target area, a common issue with sensors like Bluetooth/WiFi and RFID. While these sensors can capture transition sequence, they risk biased sampling, leading to inaccurate estimations. For example, consider a scenario where individuals with attribute 'A' have a higher rate of smartphone ownership and often opt for transition sequence 'a', while those with attribute 'B' have a lower rate of smartphone ownership and tend to prefer transition sequence 'b'. In this scenario, using Bluetooth/WiFi detection would likely result in a higher detection rate of individuals with attribute 'A' due to their higher smartphone ownership. This could lead to an overestimation of transition sequence 'a' and an underestimation of transition sequence 'b'. As this example indicates, limiting measurement scope to only a subset can lead to biased estimations.

To address the sampling bias by agent-based simulations, an effective approach is to assign measured transition sequences to agents by weighting the sequences using data from sensors capable of comprehensive measurement. Inflow count data is particularly useful for this purpose. The reason is that count data can reveal the visitation tendency of each location, allowing for an indirect estimation of the frequency of occurrence of transition sequences. Transition sequences including locations with a higher visitation rate can be assumed to have a higher frequency of occurrence, while those with a lower visitation rate can be assumed to have a lower frequency. By adjusting the frequency of occurrence of transition sequences based on the weighting of count data, it is possible to mitigate the bias caused by sampling.

\section{Experiment}
 We evaluate the efficacy of a data assimilation method that integrates agent-based simulation and particle filtering. This method is designed to tackle the three types of imperfections in people flow measurement techniques highlighted in the previous chapter. We estimate the transition tendencies of pseudo-observational data by applying a particle filter. The simulation setting involves a virtual commercial facility, which is used to simulate the shopping behavior of customers

\subsection{Model}
The proposed method can be applied to any agent-based model. For this experiment, we use a simplified version of the purchasing behavior model proposed by Hui et al(\cite{RefWorks:RefID:35-hui2009testing}). This model has been validated for its reliability based on trajectory data of thousands of customers.

The model is represented by the following formula:
\begin{eqnarray}
P_{i,j} = \frac{\exp \left[ k_i \left( A_j + \sum_{j' \neq j} \frac{A_{j'}}{(1 + d_{j,j'})^{\lambda}} \right) + \omega_i \rho_{j,t} \right]}{\sum_{j} \exp \left[ k_i \left( A_j + \sum_{j' \neq j} \frac{A_{j'}}{(1 + d_{j,j'})^{\lambda}} \right) + \omega_i \rho_{j,t} \right]}
\end{eqnarray}

where $P_{i,j}$ is the probability of agent $i$ visiting store $j$, $A_j$ is the unique attractiveness of store $j$, $d_{j,j'}$ is the distance between stores $j$ and $j'$, $\rho_{j,t}$ is the congestion level of store $j$ at time $t$, and $\omega$, $k$, $\lambda$ are coefficients. This model indicates that the probability of visiting a store is determined by its unique attractiveness, the attractiveness of surrounding stores, and the store's congestion level.

\subsection{Experimental condition}
\subsubsection{Environment}
To assess the effectiveness of data assimilation using particle filtering, we conducted tests in three distinct scenarios. These tests aim to evaluate if our method can effectively address the imperfections in people flow measurement techniques. Initially, we outline the conditions that are consistent across all scenarios.

The simulation is conducted in a virtual commercial facility with 18 stores. In the implementation, the commercial facility is represented as a graph, with each store depicted as a node. All nodes of the stores are interconnected, allowing each agent to move freely between facilities. We only record the transitions between stores, without considering the specific routes taken.

We initiated the simulation with 100 agents, assigning them random starting stores. Each agent spends 2 to 3 steps in a store before moving on to the next. After making this transition between stores three times, the agents stop roaming. Once the number of stationary agents reaches 40, we introduce an additional 40 agents to replace them. This process of replenishment continues, simulating the roaming patterns of a total of 2000 agents over 200 steps. We ran the simulations thirty times and averaged the results to obtain the final outcome.

We divided the 2000 agents into four distinct groups, with 500 in each group, all exhibiting unique behavioral patterns. These patterns are influenced by how attractive each store is. We’ve set parameter $A$ to different levels based on the store and its corresponding group: 7.5 for stores 0-2 in group 1, 8 for stores 3-5 in group 2, 8.5 for stores 6-8 in group 3, and 10 for stores 9-11 in group 4. All other stores in each group default to a value of 5. The values of the remaining parameters $\omega$, $k$, and $\lambda$ are set to 0.005, 1, and 6, respectively.

\subsubsection{Generation of pseudo observational data}
In this paper, we perform data assimilation based on pseudo-observational data generated from simulations. This approach is known as the twin experiment method, and it is an effective way to evaluate data assimilation techniques before the acquisition of real data.

To properly validate the effectiveness of the data assimilation method in twin experiments, we change the values of the simulation parameters from those used in the data assimilation environment. In this experiment, the attractiveness of each store, represented by $A$, is varied between both environments. Within the pseudo-observational environment, parameters are set as described in the previous subsection. For the data assimilation environment, all parameters are uniformly set to 5. This approach tests the efficacy of our proposed method in scenarios where information about the relative attractiveness of each store is not available.


\subsubsection{Simulation case}
The experiment is conducted under three different cases using particle filters for data assimilation:

\textit{Case 1 — Count Data Available at Each Store}

In this case, we estimate the transition tendencies from the observed data using the time-series inflow count data for each store. The sensors assumed in this setup are cameras and LiDAR. We calculate the likelihood of an agent moving to a specific store using the formula below:
\begin{eqnarray}
w^{(i)}_{\text{store}}(t) *= \exp(inflow^{(i)}_{\text{obs}}(t))
\end{eqnarray}

where $inflow^{(i)}_{\text{obs}}(t)$ represent the number of agents entering store $i$ at step $t$. 
Updating the likelihood in this way enables the simulation to accurately mirror the visiting tendencies in the real world. The number of particles is set at 100. The resampling procedure is summerized in Algorothm \ref{algprithm1}.

\begin{algorithm}[h]
\caption{Resampling process using inflow count data}
\label{algprithm1}
\begin{algorithmic}
\Procedure{Resampling process}{}
\For{$t = 1$ to $T$}
\For{$i = 1$ to $N_{store}$}
    \State $\tilde{w}_{store}^{(i)}(t) \mathrel{{*}{=}} \exp(inflow_{\text{obs}}^{(i)}(t))$
\EndFor
\State
\State $w_{store}^{(i)}(t) = \frac{\tilde{w}_{store}^{(i)}(t)}{\sum_{j=1}^{N} \tilde{w}_{store}^{(j)}(t)}$
\State
\ForAll{$i \in \text{New\_agent}$}
    \State $Store(i) = \text{select}(w_{store}(t))$
\EndFor
\EndFor
\EndProcedure
\end{algorithmic}
\end{algorithm}

\textit{Case 2 — Count Data with Attributes Available}

In this case, we estimate the transition tendencies with attribute from the observed data using the time-series attribute tagged inflow count data for each store. The sensors assumed in this setup are cameras. We calculate the likelihood of an agent moving to a specific store using the formula below: 

\begin{eqnarray}
w^{(i)}_{store}(attr, t) *= \exp(inflow^{(i)}_{\text{obs}}(attr, t))
\end{eqnarray}

where $inflow^{(i)}_{\text{obs}}(attr, t)$ represents the number of agents with the attribute $attr$ entering store $i$ at step $t$. The calculation of likelihood for each attribute follows the same formula as in Case 1. The number of particles is set at 100. The resampling procedure is summerized in Algorothm \ref{algprithm2}.

\begin{algorithm}[h]
\caption{Rresampling process using inflow count data with attribute information}
\label{algprithm2}
\begin{algorithmic}
\Procedure{Resampling process}{}
\For{$t = 1$ to $T$}
\For{attr \( = 1 \) to \( N_{attr} \)}
    \For{\( i = 1 \) to \( N_{store} \)}
        \State \( \tilde{w}_{store}^{(i)}(attr, t) \mathrel{{*}{=}} \exp(inflow_{\text{obs}}^{(i)}(attr, t)) \)
    \EndFor
    \State
    
    \State \( w_{store}^{(i)}(attr, t) = \frac{\tilde{w}_{store}^{(i)}(attr, t)}{\sum_{j=1}^{N} \tilde{w}_{store}^{(j)}(attr, t)} \)
    
    \State
    \ForAll{\( i \) in \( \text{New\_agent} \)}
        \State \( Store(attri, i) = \text{select}(w_{store}(attr, t)) \)
    \EndFor
\EndFor
\EndFor
\EndProcedure
\end{algorithmic}
\end{algorithm}

\textit{Case 3 — Count data and partial transition sequence data available}

In this case, we estimate the transition tendencies using the time-series inflow count data for each store and and biased samples of transition sequence. The sensors assumed in this setup are Bluetooth/Wi-Fi and RFID. Differing from previous cases where particles were solely model-generated, here we utilize the measured transition sequence samples as the particles. This is because using samples of real-world transition sequences allows for the replication of longer overall transition tendencies. We calculate the likelihood of the transition sequences using the formula below: 

\begin{eqnarray}
w^{(i)}_{store}(t) *= \exp(inflow^{(i)}_{\text{obs}}(t))
\end{eqnarray}
\begin{eqnarray}
w^{(i)}_{sequence}(t) = \sum_{k \in sequence(i)} w^{(k)}_{store}(t)
\end{eqnarray}

where $w^{(k)}_{store}(t)$ represents the weight of the transition sequence $k$.
This implies that sequences including stores with high inflow counts are more likely to be selected.  This likelihood calculation is similar manner to that of Nakamura et al(\cite{RefWorks:RefID:51-nakamuraestimation}). In this case, we assume that the samples are biased and consider a scenario where agents with four different attributes are sampled at respective ratios of 0.4, 0.25, 0.2, and 0.15. To evaluate the effectiveness of our proposed method under conditions of biased sampling, we compare the accuracy of selecting sequences randomly versus selecting them based on the likelihood. The number of particles is set at 400. The resampling procedure is summerized in Algorothm \ref{algprithm3}.

\begin{algorithm}
\caption{Resampling process using inflow count data and partial transition sequence data}
\label{algprithm3}
\begin{algorithmic}
\Procedure{Resampling process}{}
\For{$t = 1$ to $T$}
\State
Update store weights based on agent inflow
\For{$i = 1$ to $N_{store}$}
    \State $\tilde{w}_{store}^{(i)}(t) \gets \tilde{w}_{store}^{(i)}(t) \cdot \exp(inflow_{\text{obs}}^{(i)}(t))$
\EndFor

\State
Normalize store Weights
\State $w_{store}^{(i)}(t) \gets \frac{\tilde{w}_{store}^{(i)}(t)}{\sum_{j=1}^{N} \tilde{w}_{store}^{(j)}(t)}$

\State

Sequence weights are the total sum of store weights it passes through
\For{$i = 1$ to $N_{sequence}$}
    \State $\tilde{w}_{sequence}^{(i)}(t) \gets \sum_{k \in sequence(i)} w_{store}^{(k)}(t)$
\EndFor

\State

Normalize sequence Weights
\State $w_{sequence}^{(i)}(t) \gets \frac{\tilde{w}_{sequence}^{(i)}(t)}{\sum_{j=1}^{N} \tilde{w}_{sequence}^{(j)}(t)}$

\State
Set agent sequences based on sequence weights:
\For{$i \in \text{New\_agent}$}
    \State $Sequence^{(i)}(t) \gets \text{select}(w_{sequence}(t))$
\EndFor
\EndFor
\EndProcedure
\end{algorithmic}
\end{algorithm}

\section{Result and Discussion}
\begin{figure}[t]
\centering
\includegraphics[width=1.0\textwidth]{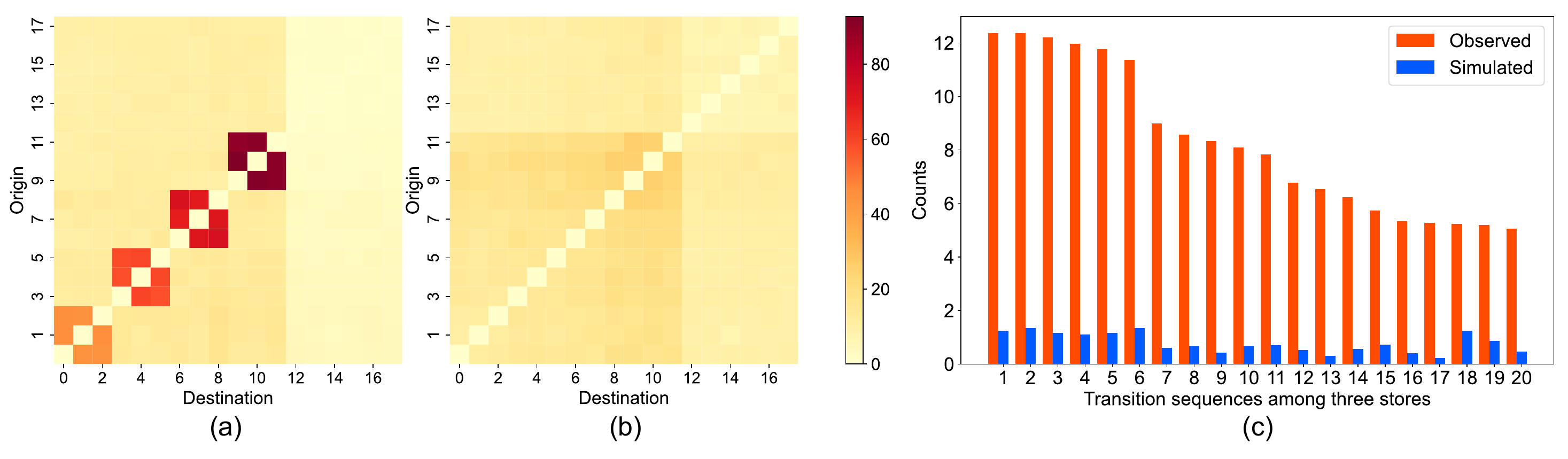}
\caption{Comparison of observational and simulated transition tendencies in case 1. (a) represents observational values of the OD (Origin-Destination) matrix, and (b) shows the estimated values based on data assimilation. The axis values represent store numbers, with the vertical axis indicating the store of departure and the horizontal axis indicating the arrival store. (c) compares the top 20 most frequent three-store transition sequences observed and estimated. The horizontal axis numbers represent the ranking of transition frequency, and the vertical axis shows their frequencies.}
\label{case1}
\end{figure}

Figure \ref{case1}(a) and \ref{case1}(b) compare the OD matrix from the observed environment with those from the data assimilation environment in Case 1. It is clear that estimating pedestrian movement tendencies between stores based solely on inflow count data is not effective. Figure \ref{case1}(c) presents a comparison of the transition tendencies among three stores, further indicating that the proposed method falls short in estimation accuracy. This difficulty arises from the uncertainty associated with inflow count data, which does not allow for individual identification. As this experiment assumes a variety of roaming behaviors, many stores exhibit high inflow counts, resulting in a wide range of transit options for them. This uncertainty leads to imprecise estimations.

\begin{figure}[!t]
\centering
\includegraphics[width=1.0\textwidth]{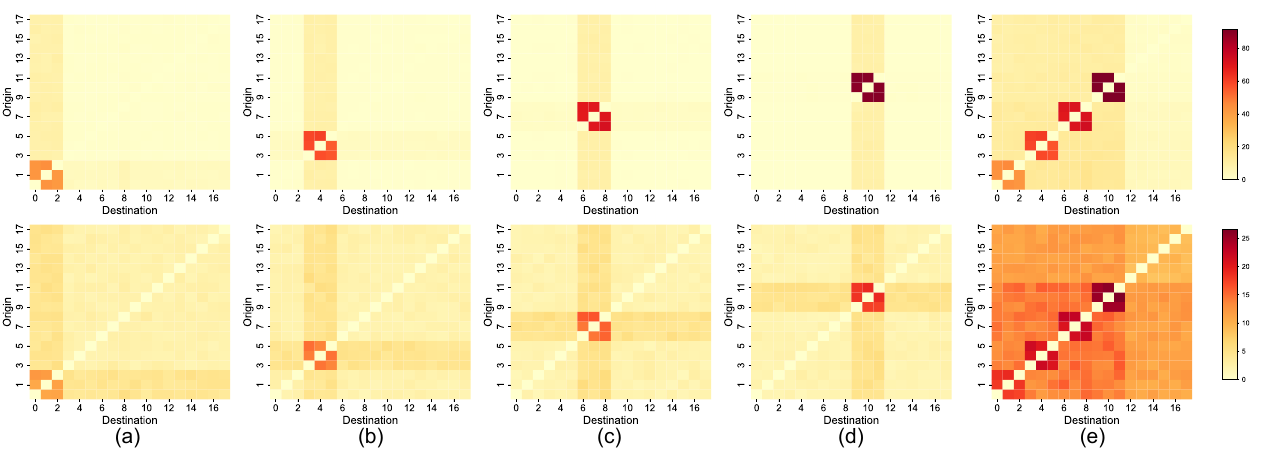}
\caption{Comparison of the observational and simulated OD matrices for each attribute (from (a) to (d)), along with the combined OD matrix (e), in Case 2. The OD matrices from (a) to (d) are derived from data assimilation, utilizing inflow count data that can distinguish between attributes.}
\label{case2}
\end{figure}

Figure \ref{case2}  compares the OD matrices from the observed environment with those from the data assimilation environment for each attribute in Case 2. The transition tendencies of each attribute are relatively well replicated. This improved accuracy results from the addition of attribute information, which reduces uncertainty of estimation. However, this scenario presumes an ideal situation where agents’ behavioral characteristics are categorized by attributes. In reality, individuals sharing the same attribute often display diverse behaviors, which could increase uncertainty. The practical effectiveness of this method in real-world observations will need to be verified in future work.

\begin{figure}[!h]
\centering
\includegraphics[width=1.0\textwidth]{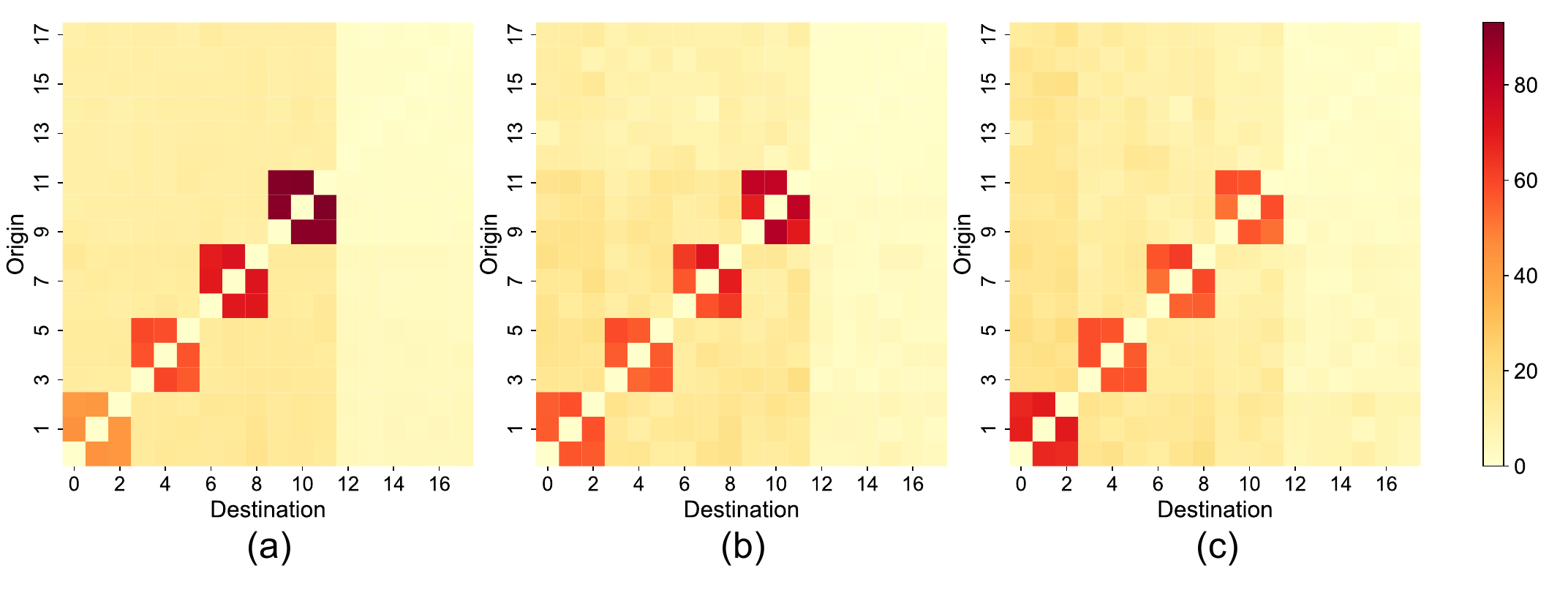}
\caption{Comparison of observational OD matrix (a) and the estimated OD matrices when data assimilation was performed (b) and not performed (c) in case 3.}
\label{case3_od}
\end{figure}

\begin{figure}[!h]
\centering
\includegraphics[width=1.0\textwidth]{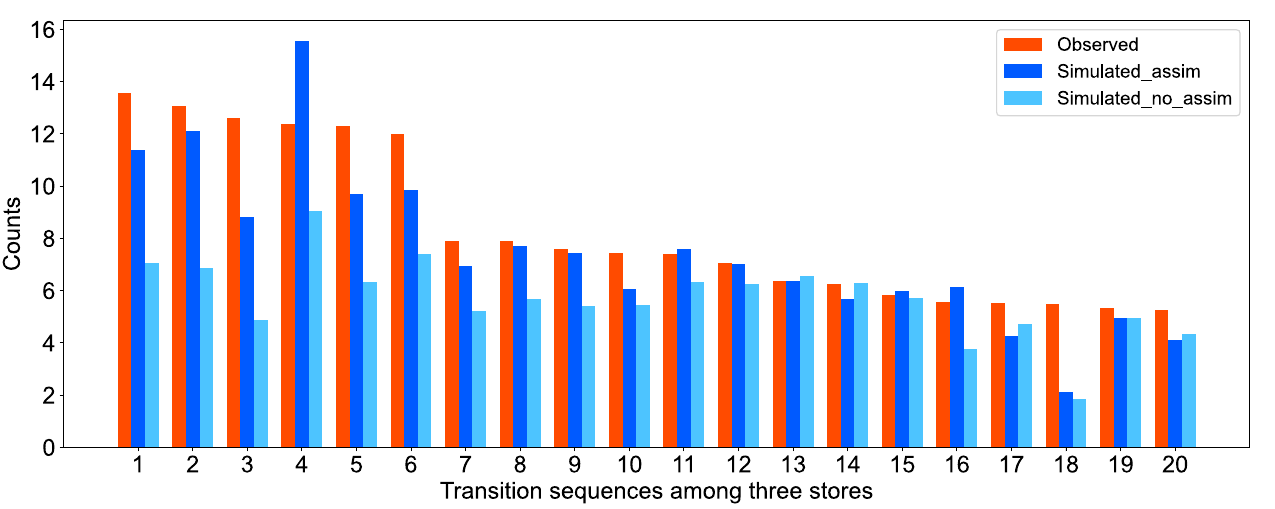}
\caption{
Comparison of the frequency of occurrence of three-store transition sequences among observational values (red), and the estimated values when data assimilation was performed (blue) and not performed (light blue) in case 3.}
\label{case3_path}
\end{figure}

Figure \ref{case3_od} compares the OD matrices from the observed environment (a) with those from the simulation environment, both with (b) and without (c) data assimilation, in Case 3. The figures demonstrate that data assimilation more effectively replicates observational transition tendencies in scenarios with biased sampling. We calculated the discrepancy between the two OD matrices, defined as the total sum of the absolute differences in the values of the matrices. The discrepancy is 715 for the data assimilation environment, compared to 1061 for the environment without data assimilation, indicating a more accurate quantitative estimation of overall transition tendencies. Figure \ref{case3_path} examines transition tendencies between three stores, revealing that the data assimilation environment also achieves relatively high estimation accuracy. These findings suggest that comprehensive measurement of inflow count data helps to mitigate the bias in sequential transition data.

The experiments in these three cases indicate the following regarding the three types of imperfections in people flow measurement methods: First, compensating for 'missing travel data' is challenging with only inflow count data; however, incorporating attribute information or partially sampled transition sequences appears effective. To address the 'lack of attribute data', using attribute-tagged inflow count data to complement transition sequences is beneficial. Lastly, to overcome the 'limited measurement scope', integrating comprehensive inflow count data proves effective in reducing sample bias.

\section{Limitation and Future Work}

\subsection{Evaluation based on actual measurement data}
In this study, the observational data utilized for data assimilation were pseudo-generated from a model and did not accurately represent real-world people flow. Therefore, it's essential to assess the system's effectiveness using actual data reflecting diverse human behaviors. 

In addition, when collecting real-world data, it's important to take into account the potential impact of measurement errors. For instance, when gathering inflow count data, sensors such as cameras and LiDAR are susceptible to errors caused by issues like occlusion. Future work should focus on determining how significantly these errors influence the proposed system.
\subsection{Verification of the dynamism of the proposed system}
The data assimilation method implemented in this study is sequential, enabling a dynamic representation of real-world people flow by progressively incorporating observational data(\cite{RefWorks:RefID:68-swarup2020live}). Understanding people flow in real-time is valuable for making decisions about congestion relief measures and digital advertising strategies. However, for real-time operation to be feasible, maintaining a low computational load is critical. Especially, the particle filter used in this research increases the computational burden as the number of particles grows, necessitating careful determination of the optimal number of particles and assimilation frequency, informed by empirical experiments.

\section{Conclusion}
This paper proposed a data assimilation framework combining agent-based simulation and particle filtering to complement the imperfection in people flow measurement techniques. We identified three main shortcomings of existing people flow measurement techniques as 'missing travel data', 'lack of attribute data', and 'limited measurement scope'. To address these imperfections, we proposed an integrative approach of measurement and simulation, using a combination of agent-based simulation and particle filter techniques for data assimilation.

As an evaluation of our proposed method, we conducted data assimilation based on pseudo-observational data and estimated transition tendencies of people in a virtual commercial facility. The measurement data were based on inflow count data from sensors installed in each store, and we tested whether the proposed method could compensate for the three identified imperfections of the measurement techniques. The results indicated that estimating transition tendencies between two points using only inflow count data is difficult due to the high level of uncertainty involved. However, the estimation accuracy improves  when attribute information is incorporated with the inflow count data. This indicates sensors such as cameras have potential usage for combining attribute information and roaming tendencies of people. Furthermore, transition tendencies between two or three points can be estimated using a combination of sample of actual transition sequence data and inflow count data, even when the samples are biased. This indicates that distributed inflow count data can modified biased sampling path data obtained from sensors such as bluetooth/wifi. The insights gained from this study are crucial for future efforts to gather people flow data, which is often challenging to capture solely through conventional measurement techniques.
\section{Acknowledgement}
This work was supported by JST SPRING, Grant Number JPMJSP2108.
\bibliographystyle{unsrtnat}  

\bibliography{template} 

\end{document}